\documentclass[trackchanges,twocolumn,twocolappendix]{aastex7}

\usepackage{graphicx}

\begin{document}

\title{Second-timescale Glints from Satellites and Space Debris Detected with Tomo-e Gozen}

\author[orcid=0000-0001-8253-6850,sname='Masaomi Tanaka']{Masaomi Tanaka}
\affiliation{Astronomical Institute, Tohoku University, Sendai, Miyagi 980-8578, Japan}
\affiliation{Division for the Establishment of Frontier Sciences, Organization for Advanced Studies, Tohoku University, Sendai 980-8577, Japan}
\affiliation{Kavli Institute for the Physics and Mathematics of the Universe (WPI), UT Institute for Advanced Study, The University of Tokyo, Kashiwa, Chiba 277-8583, Japan}
\email[show]{masaomi.tanaka@astr.tohoku.ac.jp}  

\author[]{Ichiro Takahashi}
\affiliation{Department of Physics, Institute of Science Tokyo, 2-12-1 Ookayama, Meguro-ku, Tokyo 152-8551, Japan}
\email{takahashi.i.ad@m.titech.ac.jp}

\author[orcid=0000-0001-7925-238X]{Naoki Yoshida}
\affiliation{Kavli Institute for the Physics and Mathematics of the Universe (WPI), UT Institute for Advanced Study, The University of Tokyo, Kashiwa, Chiba 277-8583, Japan}
\affiliation{RIKEN Center for Advanced Intelligence Project, 1-4-1 Nihonbashi, Chuo, Tokyo 103-0027, Japan}
\affiliation{Department of Physics, The University of Tokyo, 7-3-1 Hongo, Bunkyo, Tokyo 113-0033, Japan}
\affiliation{Research Center for the Early Universe, Graduate School of Science, The University of Tokyo, 7-3-1 Hongo, Bunkyo, Tokyo 113-0033, Japan}
\email{nyoshida94@g.ecc.u-tokyo.ac.jp}

\author[orcid=0000-0001-5701-9333]{Naonori Ueda}
\affiliation{NTT Communication Science Laboratories, 2-4 Hikaridai, Seika-cho, Keihanna Science City, Kyoto 619-0237, Japan}
\email{naonori.ueda@riken.jp}

\author[orcid=0009-0007-3042-6810]{Akisato Kimura}
\affiliation{NTT Communication Science Laboratories, 2-4 Hikaridai, Seika-cho, Keihanna Science City, Kyoto 619-0237, Japan}
\email{akisato@ieee.org}

\author[]{Kazuma Mitsuda}
\affiliation{Deloitte Tohmatsu Risk Advisory, Marunouchi, Chiyoda-ku, Tokyo 100-8360, Japan}
\email{kazuma.mitsuda@tohmatsu.co.jp}

\author[orcid=0000-0001-6020-517X]{Hirofumi Noda}
\affiliation{Astronomical Institute, Tohoku University, Aoba, Sendai 980-8578, Japan}
\email{hirofumi.noda@astr.tohoku.ac.jp}

\author[orcid=0000-0002-8792-2205]{Shigeyuki Sako}
\affiliation{Institute of Astronomy, Graduate School of Science, The University of Tokyo, 2-21-1 Osawa, Mitaka, Tokyo 181-0015, Japan}
\affiliation{UTokyo Organization for Planetary Space Science, The University of Tokyo, 7-3-1 Hongo, Bunkyo-ku, Tokyo 113-0033, Japan}
\affiliation{Collaborative Research Organization for Space Science and Technology, The University of Tokyo, 7-3-1 Hongo, Bunkyo-ku, Tokyo 113-0033, Japan}
\email{sako@ioa.s.u-tokyo.ac.jp}

\author[orcid=0000-0002-2721-7109]{Noriaki Arima}
\affiliation{Department of Aerospace Engineering, Nihon University, 7-24-1 Narashinodai, Funabashi, Chiba 274-8501, Japan}
\email{arima.noriaki@nihon-u.ac.jp}

\author[orcid=0000-0001-6402-1415]{Mitsuru Kokubo}
\affiliation{National Astronomical Observatory of Japan, National Institutes of Natural Sciences, 2-21-1 Osawa, Mitaka, Tokyo 181-8588, Japan}
\email{mitsuru.kokubo@nao.ac.jp}

\author[orcid=0000-0001-7449-4814]{Tomoki Morokuma}
\affiliation{Astronomy Research Center, Chiba Institute of Technology, 2-17-1 Tsudanuma, Narashino, Chiba 275-0016, Japan}
\email{tmorokuma@perc.it-chiba.ac.jp}

\author[]{Yuu Niino}
\affiliation{Kiso Observatory, Institute of Astronomy, Graduate School of Science, The University of Tokyo, 10762-30 Mitake, Kiso-machi, Kiso-gun, Nagano 397-0101, Japan}
\email{yuuniino@ioa.s.u-tokyo.ac.jp}

\author[orcid=0000-0001-8537-3153]{Nozomu Tominaga}
\affiliation{National Astronomical Observatory of Japan, National Institutes of Natural Sciences, 2-21-1 Osawa, Mitaka, Tokyo 181-8588, Japan}
\affiliation{Astronomical Science Program, Graduate Institute for Advanced Studies, SOKENDAI, 2-21-1 Osawa, Mitaka, Tokyo 181-8588, Japan}
\affiliation{Faculty of Science and Engineering, Department of Physics, Konan University, 8-9-1 Okamoto, Kobe, Hyogo 658-8501, Japan}
\email{nozomu.tominaga@nao.ac.jp}

\author[]{Kenzo Kinugasa}
\affiliation{Kiso Observatory, Institute of Astronomy, Graduate School of Science, The University of Tokyo, 10762-30 Mitake, Kiso-machi, Kiso-gun, Nagano 397-0101, Japan}
\email{kinugasa@ioa.s.u-tokyo.ac.jp}

\author[]{Naoto Kobayashi}
\affiliation{Kiso Observatory, Institute of Astronomy, Graduate School of Science, The University of Tokyo, 10762-30 Mitake, Kiso-machi, Kiso-gun, Nagano 397-0101, Japan}
\email{naoto@ioa.s.u-tokyo.ac.jp}

\author[]{Sohei Kondo}
\affiliation{Kiso Observatory, Institute of Astronomy, Graduate School of Science, The University of Tokyo, 10762-30 Mitake, Kiso-machi, Kiso-gun, Nagano 397-0101, Japan}
\email{kondo@ioa.s.u-tokyo.ac.jp}

\author[]{Yuki Mori}
\affiliation{Kiso Observatory, Institute of Astronomy, Graduate School of Science, The University of Tokyo, 10762-30 Mitake, Kiso-machi, Kiso-gun, Nagano 397-0101, Japan}
\email{moriyuki@ioa.s.u-tokyo.ac.jp}

\author[orcid=0000-0001-5797-6010]{Ryou Ohsawa}
\affiliation{National Astronomical Observatory of Japan, National Institutes of Natural Sciences, 2-21-1 Osawa, Mitaka, Tokyo 181-8588, Japan}
\email{ryou.ohsawa@nao.ac.jp}

\author[]{Hidenori Takahashi}
\affiliation{Kiso Observatory, Institute of Astronomy, Graduate School of Science, The University of Tokyo, 10762-30 Mitake, Kiso-machi, Kiso-gun, Nagano 397-0101, Japan}
\email{nori@ioa.s.u-tokyo.ac.jp}

\author[orcid=0009-0007-4821-5827]{Satoshi Takita}
\affiliation{Institute of Astronomy, Graduate School of Science, The University of Tokyo, 2-21-1 Osawa, Mitaka, Tokyo 181-0015, Japan}
\email{takita@ioa.s.u-tokyo.ac.jp}



\begin{abstract}
A search for second-timescale optical transients is one of the frontiers
of time-domain astronomy.
However, it has been pointed out that reflections of sunlight
from Earth-orbiting objects can also produce second-timescale ``glints.''
We conducted wide-field observations at 2 frames per second
using Tomo-e Gozen on the 1.05 m Kiso Schmidt telescope.
We identified 1554 point-source glints that appeared in only one
frame (0.5 sec).
Their brightness ranges from 11 to 16 mag, with fainter glints being more numerous.
These glints are likely caused by satellites and space debris
in high-altitude orbits such as the geosynchronous Earth orbit
or highly elliptical orbits.
Many glints brighter than 14 mag are associated with known
satellites or debris with large apogees ($>$30,000 km).
In contrast, most fainter glints are not associated with
cataloged objects and may be due to debris with sizes of 0.3--1 m.
The event rate of second-timescale glints is estimated to be 
$4.7 \pm 0.2\ {\rm deg^{-2}\ hr^{-1}}$ (average)
and $9.0 \pm 0.3\ {\rm deg^{-2}\ hr^{-1}}$ (near the equator) at 15.5 mag.
Our results demonstrate that high-altitude debris
can represent a significant foreground in searches for second-timescale optical transients.
They also imply that deep surveys such as Rubin/LSST 
will detect many of these glints in single-exposure images.  
\end{abstract}



\begin{deluxetable*}{lllllll}
\tablewidth{0pt}
\tablecaption{Summary of second-timescale glints \label{tab:rate}}
\tablehead{
  \colhead{Telescope/Camera} & \colhead{FOV}        & \colhead{Exp. time} & \colhead{Magnitude} &  \colhead{Magnitude} & \colhead{Event rate}                   & \colhead{References}\\
                      & \colhead{(deg$^2$)}   & \colhead{(sec)}     &                    &   \colhead{(0.5 sec)} & \colhead{(${\rm deg^{-2}\ hr^{-1}}$)}  & \colhead{}
}
\startdata
Evryscope         & 16,512  & 120   & 14    &  8    & $0.043^{+0.0067}_{-0.014}$$^a$ & \citet{corbett20}  \\
W-FAST            & 7       & 0.04  & 9--11  & 10--12 $^{\dagger}$ & $1.74^{+0.16}_{-0.17}$$^b$     & \citet{nir21}  \\
                  &         &       &       &       & $1.26^{+0.16}_{-0.17}$$^c$     &                  \\
ZTF               & 47      & 30    &       &      13.8$^{\dagger\dagger}$ & 1.0--1.9$^a$               & \citet{karpov23}$^{\dagger\dagger\dagger}$ \\
Tomo-e Gozen      & 20      & 0.5   & 15.5  &  15.5   & $4.7\pm 0.2^a$               & This work \\
                  &         &       &       &         & $9.0\pm 0.4^d$               & \\
                  &         &       &       &         & $2.1\pm 0.1^e$               & 
\enddata
\tablecomments{All the event rates are given in a detection-wise mannar (not object-wise). \\
  $^{\dagger}$ Converted to 0.5 sec equivalent magnitude by assuming the intrinsic duration of $\Delta t=0.2$ sec.\\
$^{\dagger\dagger}$ Converted to 0.5 sec equivalent magnitude from the typical instantaneous magnitude of 12 mag in the duration of $\Delta t = 0.1 $ sec. \\
  $^{\dagger\dagger\dagger}$ The lower value counts only catalogued objects (including those at the LEO) while the upper value also counts uncatalogued objects. \\
  $^a$ Average, $^b$ Dec around $-25$ deg, $^c$ Dec around $+5$ deg, $^d$ Equator ($-20$ deg $<$ Dec $<$ +20 deg), $^e$ $|{\rm Dec}| >$ 20 deg.}
\end{deluxetable*}

\section{Introduction} 
\label{sec:intro}

Significant progress in optical time-domain survey
over the last few decades has enabled studies of the dynamic Universe.
In particular, transient phenomena with timescales
longer than days have been extensively studied by 
time-domain surveys such as the Palomar Transient Factory \citep{rau09},
ASAS-SN \citep{shappee14,kochanek17}, Pan-STARRS1 (PS1, \citealt{chambers16}),
ATLAS \citep{tonry18}, and the Zwicky Transient Facility (ZTF, \citealt{bellm19}).
Also, some surveys have explored transient sky
with a shorter timescale, such as hours 
\citep[e.g.,][]{becker04,rau08,lipunov07,ho18,vanroestel19,oshikiri24}
or even minutes \citep[e.g.,][]{berger13fot,andreoni20}.

One of the frontiers of time-domain survey is the exploration of the
optical transient sky with a timescale of seconds or even shorter.
Such transient phenomena are known in other electromagnetic wavelengths
such as gamma-ray bursts (GRB, second timescale, \citealt{woosley06})
and fast radio bursts (FRB, millisecond timescale, \citealt{petroff22}).
However, such astrophysical transients have never been
identified at optical wavelengths,
except for simultaneous detection of GRBs (e.g., \citealt{racusin08}).
Nevertheless, some works have begun to place constraints
on the event rate of second-timescale optical transients 
\citep{richmond20,arimatsu21,nir21,arima25}.

A big challenge in the search for second-timescale optical transients
is the presence of large numbers of foreground objects
caused by satellites or space debris
orbiting the Earth and reflecting sunlight.
In fact, a satellite body passing the spectroscopic slit
was mistaken for a high-redshift GRB
  for the case of GN-z11 \citep{jiang21,nir21b,michalowski21}.
In addition, when these objects reflect the sunlight only for
a short timescale,
they can even produce short-timescale flashes.
Such satellite ``glints'' were already recognized in the 1980s
\citep{katz86,maley87,schaefer87}.
And the number of satellites and space debris has steadily increased
since then, making the event rate of glints higher.

Recent systematic time-domain surveys actually detect
many short-timescale glints.
For example, \citet{corbett20} identified a large number of satellite glints
with Evryscope.
\citet{karpov23} also detected many glints in the ZTF data.
They intensively crossmatched the detected sources, and the orbit of the
detected objects range from the Low Earth Orbit (LEO, altitude of $\le$ 2000 km)
to the Geosynchronous Earth Orbit (GEO, $\sim$ 36,000 km).

These works selected objects whose shape is consistent
with a point spread function (PSF) of surrounding stars.
For moving objects to keep the point source shape within their exposure time,
their intrinsic duration of the flash should be significantly shorter
than the exposure time.
For the case of glints detected with ZTF (30 sec exposure, \citealt{karpov23}),
the expected duration is an order of $10^{-3}$ sec at LEO
(typical apparent motion is $\sim 240$ arcsec/sec)
and $10^{-1}$ sec at the GEO ($\sim 15$ arcsec/sec).

In fact, the short intrinsic duration has been directly confirmed
by time-resolved observations.
\citet{karpov16} reported detection of glints
with a timescale of 0.5 sec by resolving their brightness evolution
with 0.1 sec exposure images of Mini-Mega-TORTORA.
Also, \citet{nir21} performed a systematic survey with
a wide-field CMOS camera on W-FAST
(The Weizmann Fast Astronomical Survey Telescope)
with 0.04 sec exposure (frame rate of 25 Hz).
They also identified a number of glints with 9--11 mag
with a timescale of about 0.1-0.3 sec.

In time-resolved observations, these glints do not significantly move
within the duration of the flash,
implying that they are caused by objects at the GEO.
In fact, population of satellites and space debris at the GEO has
been intensively studied by dedicated surveys
(see e.g., \citealt{schildknecht07} for a review).
Such survey uncovered a large number of objects with even fainter
optical magnitude (15--20 mag, e.g., \citealt{schildknecht01,jarvis01,seitzer04,barker05,molotov09,bolden11,seitzer16,blake21}).
These dedicated surveys often search for objects by stopping the telescopes,
  i.e., an object at the GEO appear as a static source.
By assuming a constant brightness, 15--20 mag at the GEO
corresponds to 0.1--1 m size objects.
Thus, we also expect fainter short-timescale glints.
However, as GEO surveys usually aims at understanding the number of
orbiting objects,
an event rate of short-timescale glints caused by
these small objects has not been extensively studied.
Table \ref{tab:rate} summarizes past estimates for the event rates
of short-timescale glints.

In this paper, we present results of our survey with
a rate of 2 frames per second (0.5 sec exposure time) with Tomo-e Gozen mounted on 1.05 m Schmidt telescope.
Our survey enables to explore second-timescale glints down to about 15--16 magnitude in 0.5 sec,
which significantly extends the searching sensitivity in this timescale
(Table \ref{tab:rate}).
In Section \ref{sec:obs}, we present our observations.
In Section \ref{sec:selection}, we describe our selection methods for
second-timescale glints.
We show our results in Section \ref{sec:results},
and discuss the nature of the glints, their event rate, and
implications for future surveys in Section \ref{sec:discussion}.
Finally, we give summary in Section \ref{sec:summary}.


\begin{figure}
  \includegraphics[width=\columnwidth]{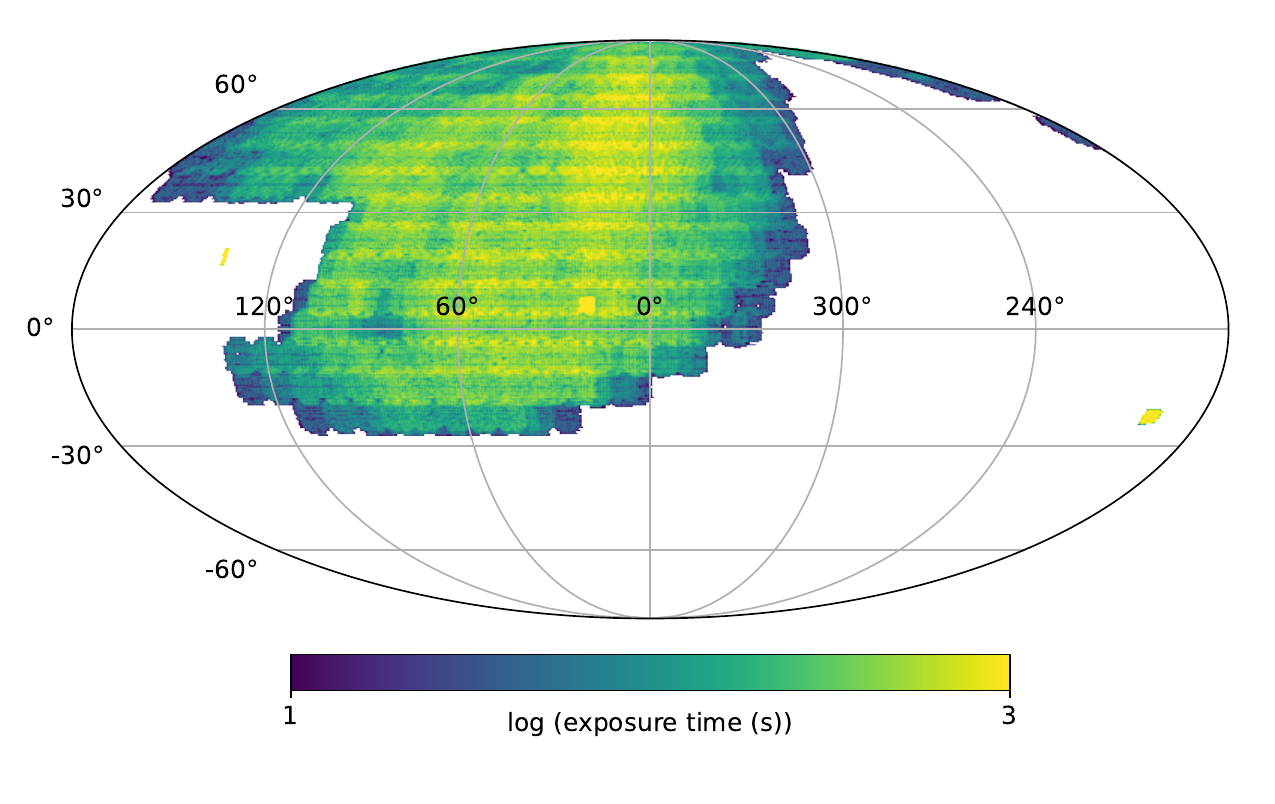}  
  \includegraphics[width=\columnwidth]{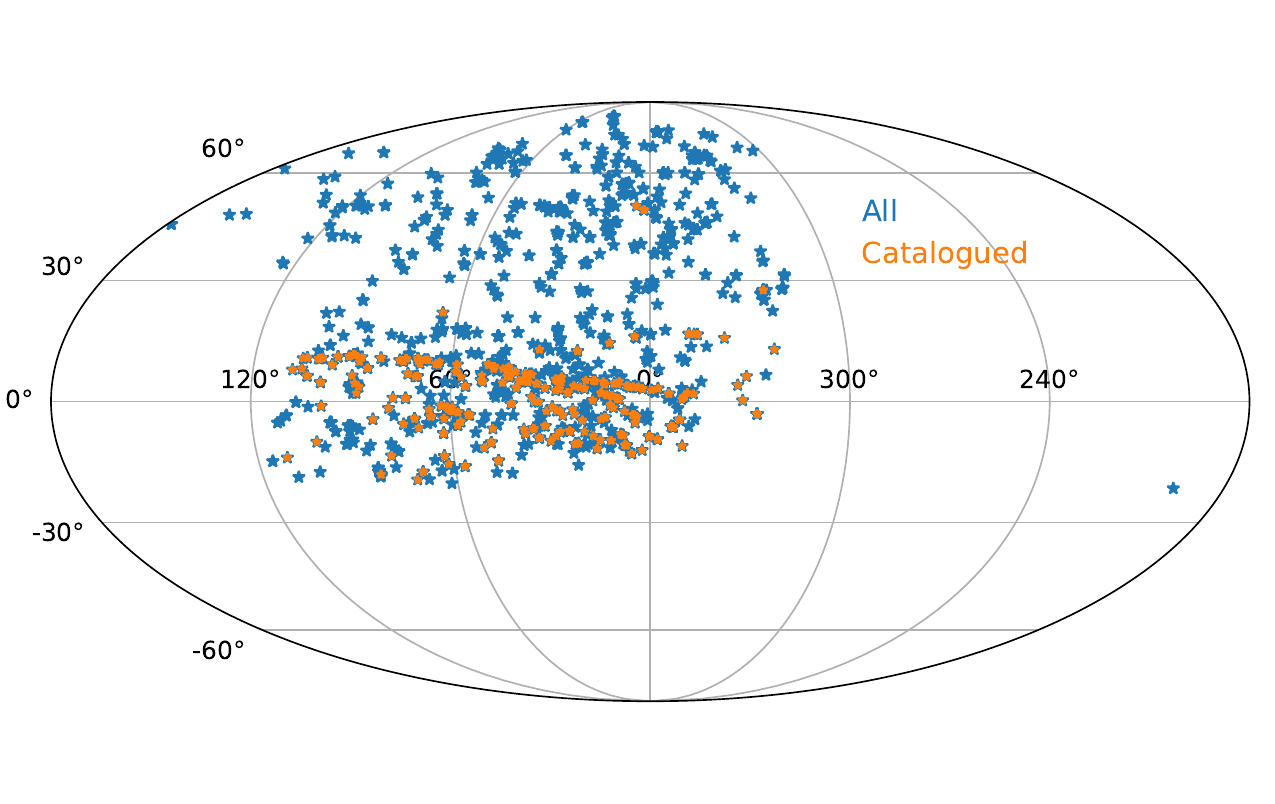}  
  \caption{(Top) Survey footprints of our observations in the equatorial coordinate. The color represents the total exposure time for each part of the sky. (Bottom) Distribution of the detected glints. The blue points show all the glints while the orange points show those associated with catalogued objects. Note that the coordinates of the glints are given as viewed from the Kiso observatory.}
    \label{fig:survey}
\end{figure}

\begin{figure}
	\includegraphics[width=\columnwidth]{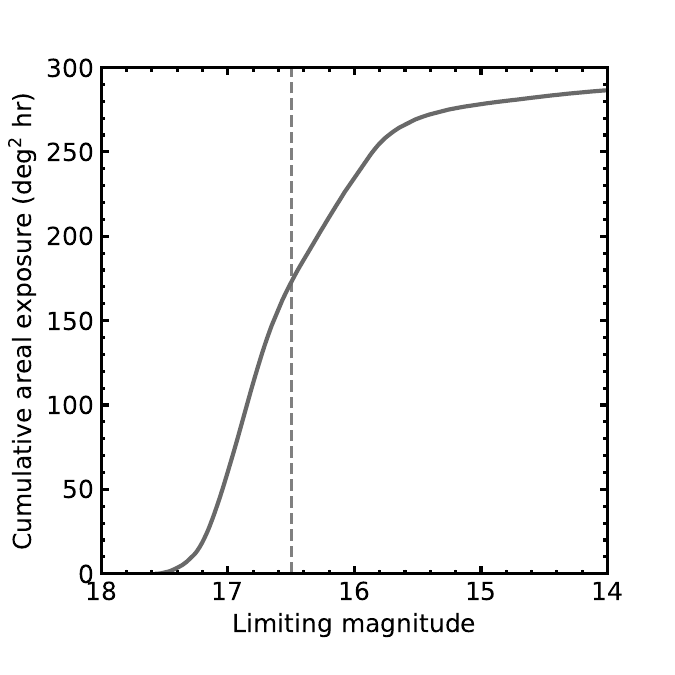}
    \caption{Cumulative areal exposure (${\rm deg^2 \ hr}$) of our survey as a function of $5 \sigma$ limiting magnitude.}
    \label{fig:areal_exposure}
\end{figure}

\section{Observations}
\label{sec:obs}

We conduct video-mode observations using Tomo-e Gozen 
\citep{sako18} mounted on the Kiso 1.05m Schmidt telescope
\footnote{The observatory is located at a longitude of 137d37m31.5s (east), 
a latitude of 35d47m50.0s (north), and an altitude of 1132 m.}.
Tomo-e Gozen is a wide-field camera with 84 CMOS sensors.
Each sensor has 2000 × 1128 pixels with a pixel scale of 1.189 arcsec
per pixel, covering an area of 0.246 deg$^2$ per sensor.
The total field of view of the camera is 20.7 deg$^2$.
The data are taken without a filter.
Photoelectric conversion efficiency of the CMOS sensor has a peak around
5000 ${\rm \AA}$, and the efficiency drops to half of the peak at 3800 and 7100 ${\rm \AA}$ \citep{kojima18}.

We use the data from 21-night observations during 2023 January to February.
For each night, about three hours after the sunset were used
to cover a large area of the observable sky.
Each visit to the sky consists of 18 consecutive frames
of 0.5 sec exposure images (2 frames per second),
which enables us to search for objects that appear only for 0.5 sec.
Each patch of the sky is visited with 4-point dithering to fill the chip gaps.
A typical $5 \sigma$ limiting magnitude (in 0.5 sec image) is 16.5 mag.

Figure \ref{fig:survey} shows our survey footprints.
The color in the figure represents the total exposure time
for each patch of the sky.
The total exposure of our survey observations can be expressed as
so-called areal exposure, a product of survey area
and time spent to monitor the area.
Figure \ref{fig:areal_exposure} shows the cumulative areal exposure
of our survey as a function of the limiting magnitude.
The total areal exposure of our survey is 290 deg$^2$ hr.
If we restrict the data to limiting magnitude deeper than 16.5 mag,
the areal exposure is 173 deg$^2$ hr.

All the imaging data are reduced by subtracting bias and dark
frames, and by correcting with flat-field frames.
Then, astrometric solution is obtained for each chip
with astrometry.net \citep{lang10}.
Source detection is performed by {\tt SExtractor} \citep{bertin96}.
By using the detected point sources,
photometric calibration is done with respect to the Gaia catalog
($G$ mag, \citealt{gaia21}).
Then, a limiting magnitude is evaluated for each chip
based on the photometric error of the detected point sources.
Detection of short-timescale glints is performed separately
as described in Section \ref{sec:selection}.

\section{Selection of second-timescale glints}
\label{sec:selection}

\subsection{Detection methods}

We have developed a machine-learning-based detection algorithm
to efficiently and rapidly detect objects that appear only in one frame (0.5 sec)
from our movie data.
Our Tomo-e data are taken at 2 frames per second with 84 CMOS sensors,
yielding the data acquisition rate of about 2 TB per hour.
As it is not feasible to keep all the movie data for a long-term survey,
transient objects need to be identified in real time
before the data are discarded, making a fast detection algorithm essential.
The total data size used in this paper is 84.7 TB (a small subset of the long-term survey),
which is entirely stored for data characterization.

For this purpose, we adopt Single Shot MultiBox Detector (SSD, \citealt{liu15}).
SSD is a method that performs object detection and classification simultaneously
using a deep neural network.
The inputs to SSD are two-dimensional images
and the outputs are the positions of the detected objects (bounding boxes)
and their classification probabilities.
We have developed dedicated software to classify the objects into
three classes: (1) stationary objects (or ``star''), (2) transients that appear
only in one frame, and (3) background (i.e., no object).
The software runs on GPUs, which enables real-time data processing.

For the input images, we use three consecutive frames
(2000 $\times$ 1128 pixels $\times$ 3 frames).
For each movie data set (18 frames), detection and classification
are repeated for 16 times by sequentially inputting three-frame sets.
Note that our method is not sensitive to transients that appear only
in the first or last frame.
Thus, the effective time is 16 $\times$ 0.5 sec = 8 sec
per 18-frame data.

The SSD model consists of a base network (as a feature extractors)
and additional convolutional layers to predict the bounding box and class.
Each pixel in the additional convolutional layers has
predefined bounding boxes, providing predictions of the position and class.
Finally, the detections with different spatial scales that yield high scores
are merged by non-maximum suppression.
The training data for the detection and classification were prepared
by using actual Tomo-e Gozen data.
As for transient category, we created training data
by injecting artificial point sources into the second frame.
More details of our detection algorithm are given in the Appendix.

Figure \ref{fig:tpr} shows the true positive rate (TPR)
of the transient detection
as a function of magnitude relative to the 5$\sigma$ limiting magnitude.
The black and blue lines show TPRs 
with a score threshold of $> 0.35$ and $> 0.7$, respectively.
We set our criterion for the score of $>0.70$ (blue line),
which gives a TPR of about 0.9 at brighter magnitudes
and a TPR of $\simeq 0.5$ at 1 mag brighter than the limiting magnitude.

\begin{figure}
	\includegraphics[width=\columnwidth]{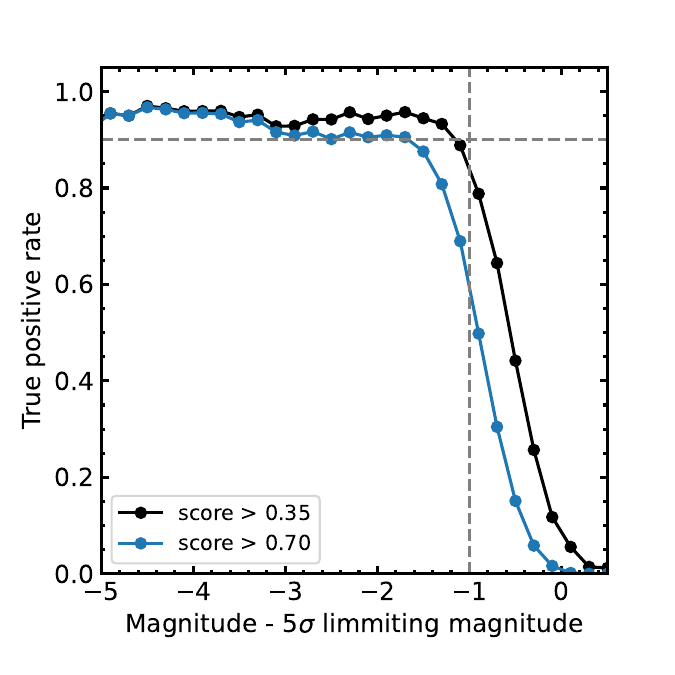}  
    \caption{TPR (or recall) as a function of magnitude relative to 5$\sigma$ limiting magnitude. Black and blue lines show TPR with score $> 0.35$ and $> 0.70$, respectively. In our analysis, we adopt the threshold score of 0.70 (blue).}
    \label{fig:tpr}
\end{figure}

\begin{figure*}
\begin{center}  
  \includegraphics[width=14cm]{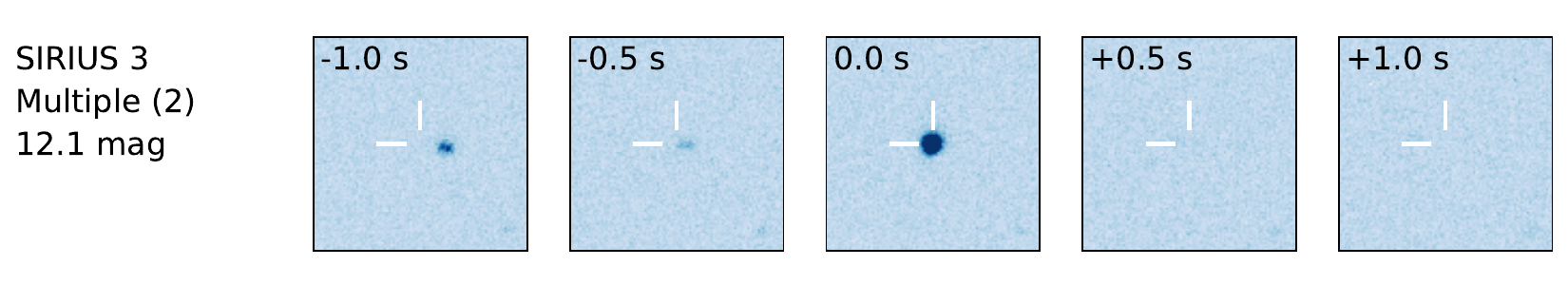}
  \includegraphics[width=14cm]{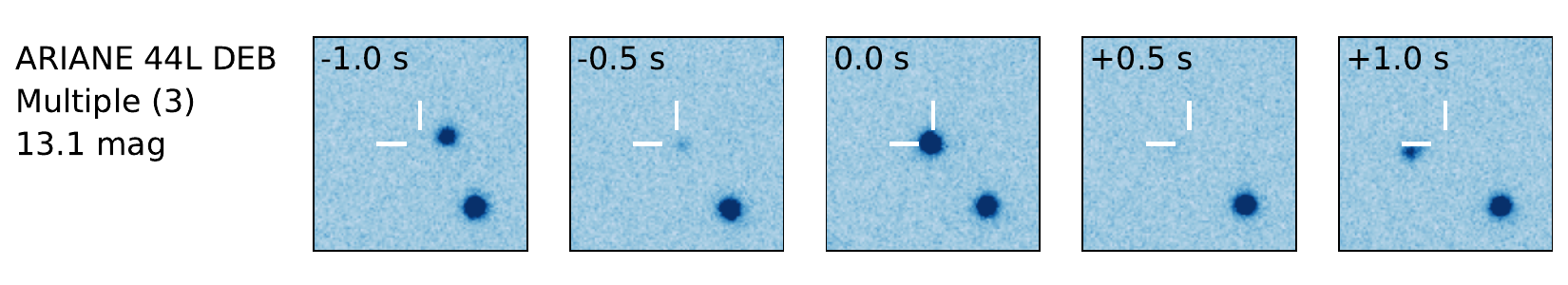}
  \includegraphics[width=14cm]{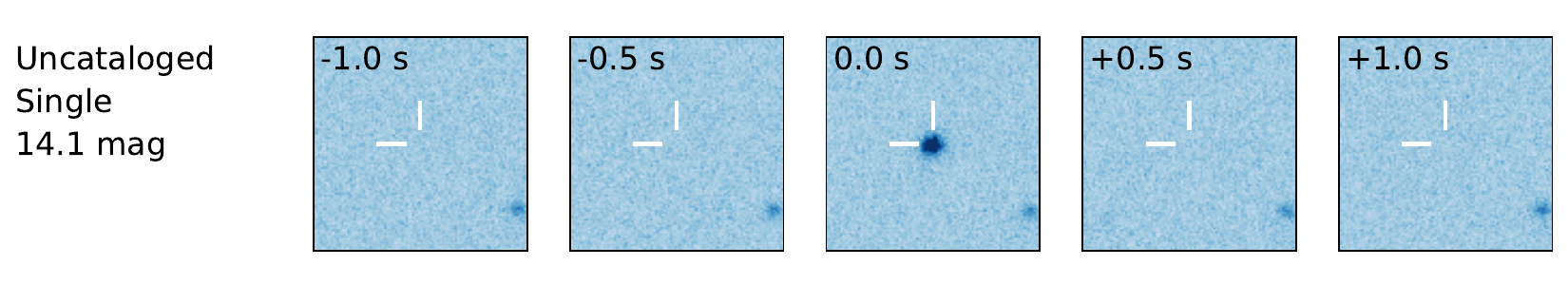}
  \includegraphics[width=14cm]{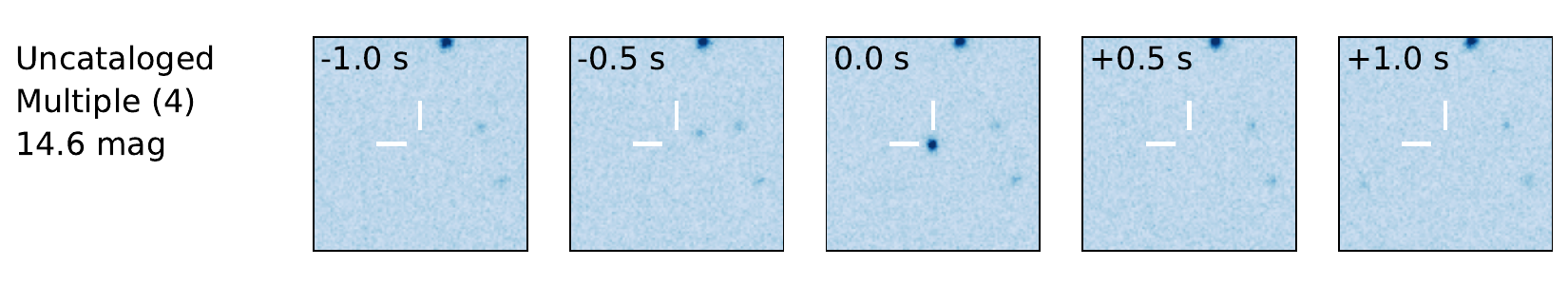}
  \includegraphics[width=14cm]{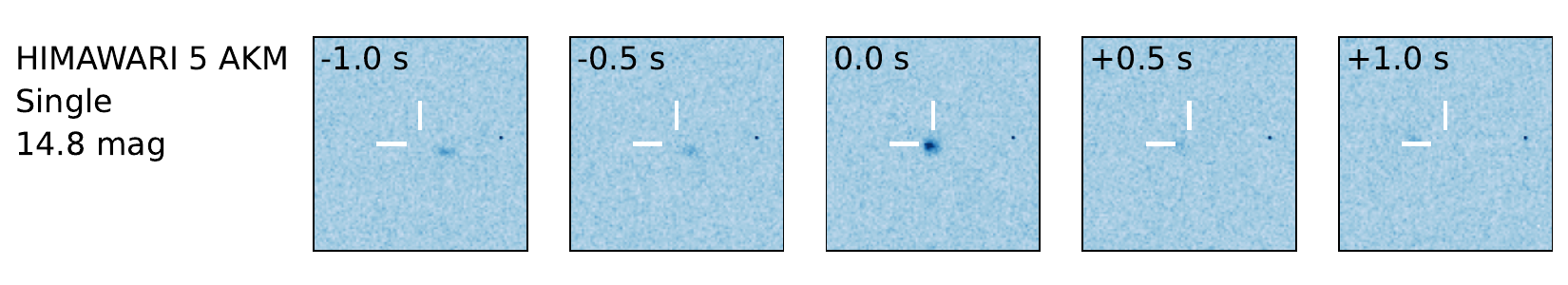}
  \includegraphics[width=14cm]{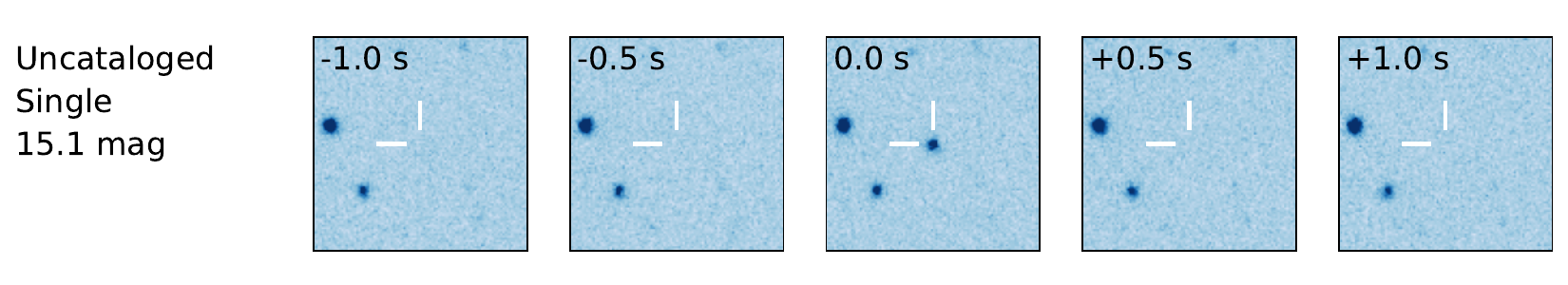}
  \includegraphics[width=14cm]{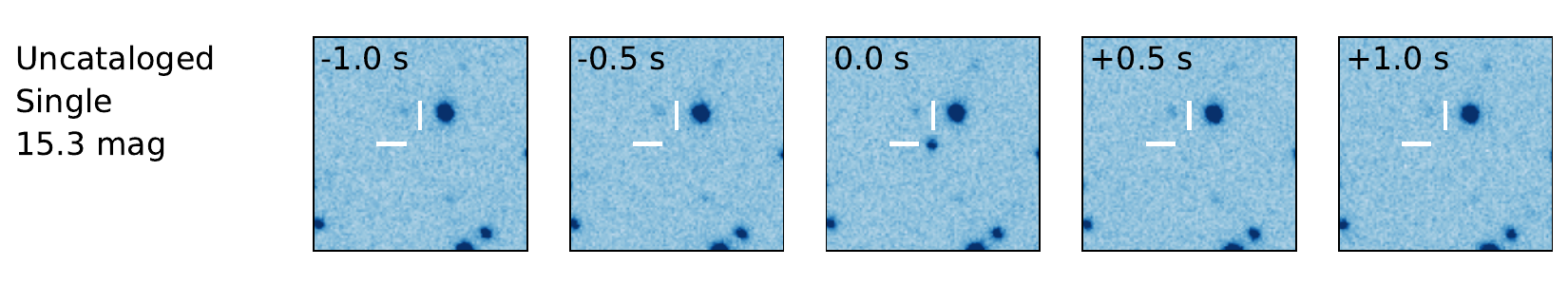}
  \includegraphics[width=14cm]{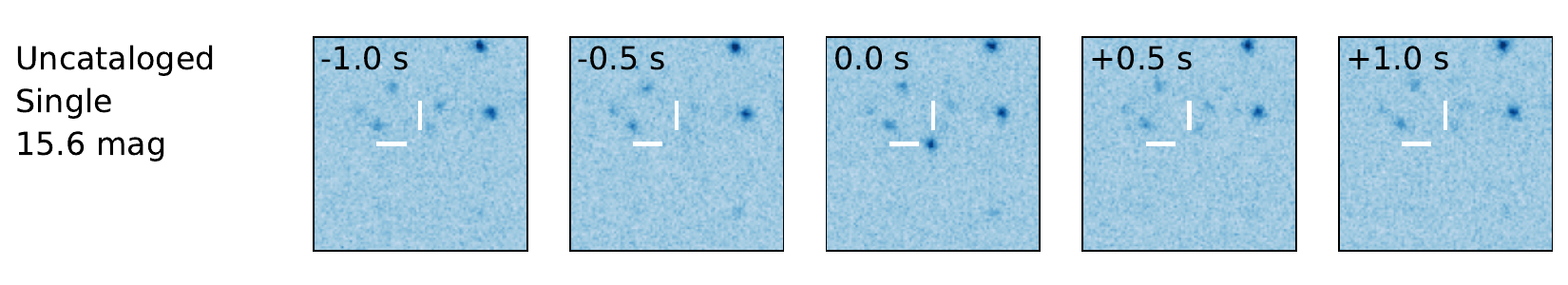}  
  \caption{Example of detected glints. For the glints matched with catalogued objects, their common names are also given. Each image section has a size of 100 pixels $\times$ 100 pixels (about 2 arcmin $\times$ 2 arcmin). North is up and East is left.
  Note that some faint sources (i.e., those in SIRIUS 3 (first row), uncatalogued source (fourth row), and HIMAWARI 5 AKM (fifth row)) are not detected with SSD, and these are not included in the number of detection.}
    \label{fig:image}
\end{center}  
\end{figure*}

\subsection{Candidate selection}

The purpose of our study is to understand the event rate of
second-timescale glints consistent with point sources
\footnote{Genuine point sources such as stars form a well-defined
PSF even in a 0.5 sec image under the observing conditions
at the Kiso observatory.}.
Although our SSD is trained to detect point sources,
the detected candidates still include elongated sources and streaks,
as there is no appropriate class for these shapes.
Thus, to remove these non-point-source objects,
we cross-match our candidates with detection catalogs
from {\tt SExtractor} \citep{bertin96}
by imposing criteria of (1) 0.7 $<$ FWHM/FWHM$_{\rm im} <$ 1.4
and (2) elon $<$ 1.4.
Here, FWHM and elon are full-width half maximum and elongation
(major/minor axis ratio)
of the detected source, respectively (FWHM$_{\rm im}$ is
the median FWHM of the detected stars).
Note that these criteria are somewhat arbitrary, leaving uncertainties in
comparisons with samples from other works
(see Section \ref{sec:discussion}).
As a result of crossmatch, 1759 point source candidates were selected.
Then, we visually inspected all of these candidates,
and removed spurious detections, such as cosmic-ray hits
and pixels around bright stars.
Finally, 1554 glints remain as the final sample.

Among the final samples, 1088 glints are identified multiple times in the same chip
at different positions in 18 time frames
(hereafter we label these as ``multiple'' glints).
Among 1088 glints with multiple detections,
312, 276, 152, and 348 glints are detected 2, 3, 4, and more than 5 times,
respectively.
The motions of the glints with more than 3 detections
can be traced with straight lines.
We here include objects with 2 detections in the ``multiple''
category as the probability of detecting two unrelated glints
in the same chip
in 9 sec of movie data is small (see Section \ref{sec:discussion} for the rate).
If we exclude objects with two detections, the number of multiple glints
is reduced by about 30\%.
The remaining 466 glints are identified only once (``single'').
Note that we do not attempt to associate the same objects appearing
in different chips.
Therefore, the number of glints above includes duplications of the same objects.

For all the detected glints, we searched for catalogued satellites or debris
using {\tt Skyfield} software \citep{rhodes19}.
We used Two-line Elements (TLE) provided by Space-Track
\footnote{\url{https://www.space-track.org}} published within 2 days of the observing date.
We regard a glint as associated if
if a known satellite (or debris) is found within 60 arcsec.
The number of glints associated with known objects is 563 in total:
  101 for single glints (22\% of all 466 single glints)
  and 462 for multiple glints (42\% of all 1088 multiple glints).
The number of satellites/debris associated with these glints is only 70 in total;
these objects appear multiple times in different chips.

A similar catalog match was also performed by \citet{karpov23}.
  They carried out an extensive search for satellite glints
  with 30 sec exposure images from ZTF.
  They identified 116,389 ``tracklet'' candidates (appearing more than five times in a straight line)
  and 151,046 ``morphology''-selected candidates (showing multiple peaks in 60 $\times $ 60 arcsec cutout images).
  Among these candidates, 74,619 (64\%) and 75,280 (50\%) candidates were associated
  with known satellites/debris, respectively.
  Our finding is consistent with their results in that a large fraction of multiple glints
  are associated with catalogued objects.
  Our fraction is somewhat smaller (42\%) most likely because
  the glints detected in our surveys tend to be fainter
  (see Sections \ref{sec:results} and \ref{sec:discussion}).

Figure \ref{fig:image} shows examples of the detected glints.
Each row shows 5 consecutive images centered on the detected position
  in the third frame.
For the glints associated with catalogued objects,
we also give their common names.

\begin{figure}
	\includegraphics[width=\columnwidth]{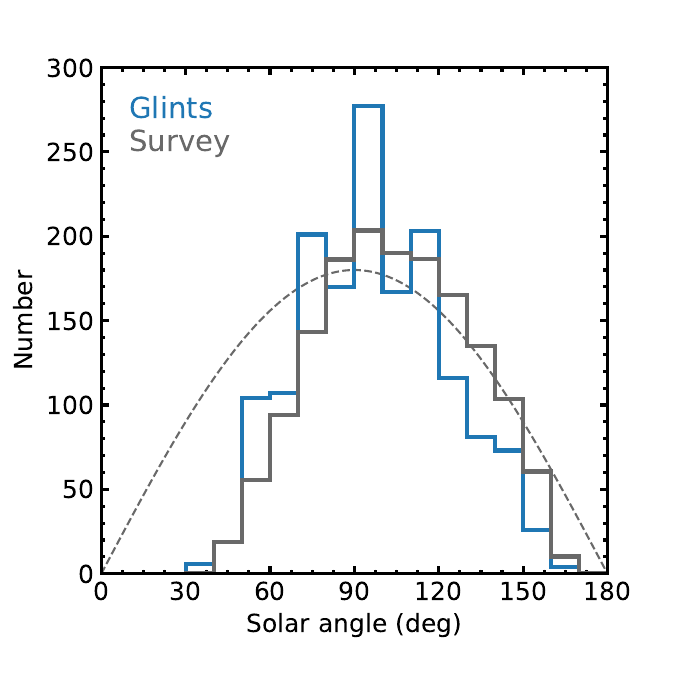}
        \caption{Distribution of solar separation angles for the detected glints (blue).
          For comparison, the gray line shows the (normalized) angle distribution for our survey footprints. The thin dashed line shows the uniform distribution.}
    \label{fig:dist_sep}
\end{figure}

\begin{figure}
	\includegraphics[width=\columnwidth]{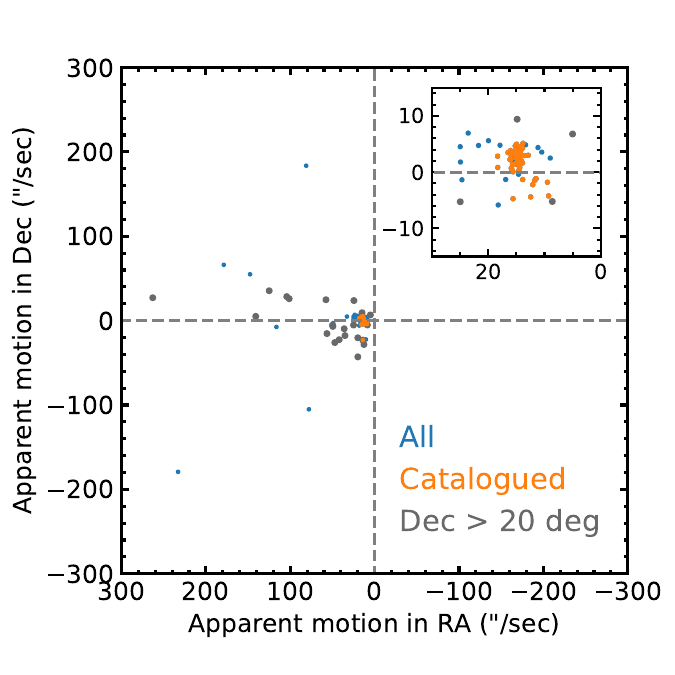}
    \caption{Apparent motion in RA and Dec for the glints detected with $\ge 3$ times (blue). The orange dots show the motions of the objects associated with catalogued objects. The gray dots show the motions of the objects at Dec $>$ 20 deg.}
    \label{fig:motion}
\end{figure}

\section{Results}
\label{sec:results}

The sky distribution of the detected glints is broadly similar to our survey footprints
with an enhancement near the equatorial plane.
The bottom panel of Figure \ref{fig:survey} shows the sky distribution of
the detected glints (blue).
It is clear that glints associated with catalogued objects are
preferentially located near the equatorial plane (orange).
Figure \ref{fig:dist_sep} shows the distribution of the solar separation angle
for the detected glints.
Their distribution has a peak around 90 deg,
which also follows the survey footprints (gray line).

For glints detected multiple times,
  we measure their apparent velocity on the sky.
  For these measurements, we only use glints with more than 3 detections
  whose motion is well described by a straight line.
Figure \ref{fig:motion} shows the apparent velocities in RA and Dec.
Most of the measured velocities are 
clustered around an RA motion of 15 arcsec/sec,
which is expected for objects at the GEO.
In fact, among the 70 catalogued objects producing glints,
59 objects have an apogee of the orbit greater than 30,000 km.
Among these 59 objects, only six objects have an eccentric orbit with a apogee $>$ 30,000 km and a perigee $<$ 5,000 km.
These results suggest that majority of the glints detected near the equator
are caused by objects at the GEO.

It is interesting that our observations also detect 
glints even at high latitudes (Figure \ref{fig:survey}).
The glints at high latitudes show a larger scatter
in their apparent velocities (blue points in Figure \ref{fig:motion}).
However, many of these objects show the apparent velocities lower than those
expected for objects at the LEO.
Thus, these glints may be caused by objects in orbits
with high altitude (apogees) and high inclinations,
such as highly elliptical orbit (HEO).

\begin{figure}
  \includegraphics[width=\columnwidth]{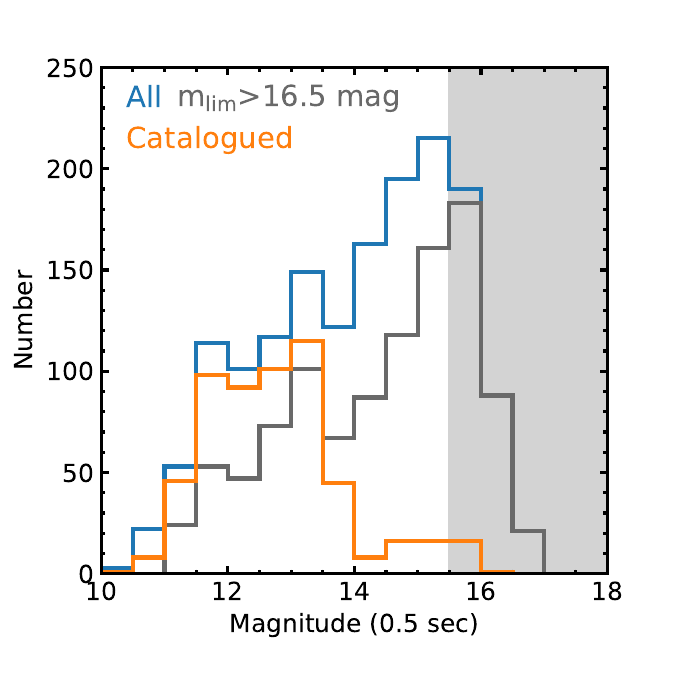}
  \includegraphics[width=\columnwidth]{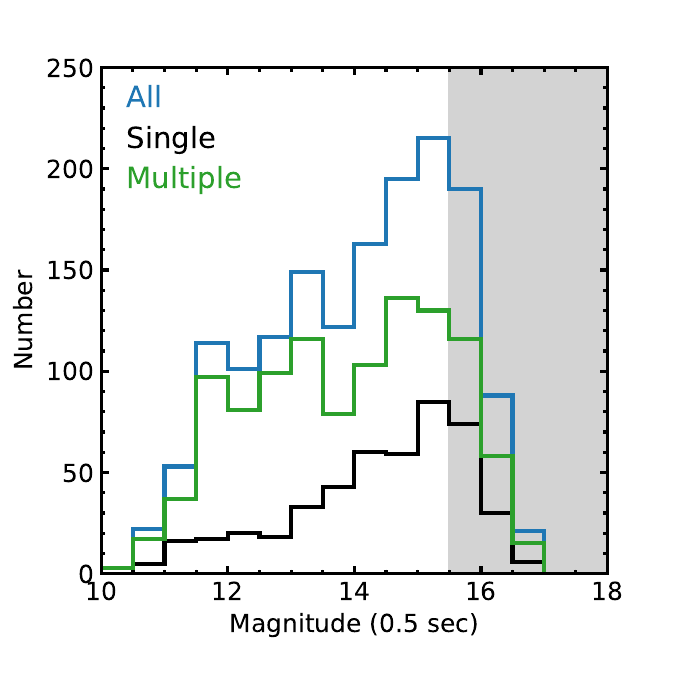}  
  \caption{Magnitude distribution of the detected glints (blue). (Top) Gray line shows the magnitudes of the glints detected in images deeper than 16.5 mag, which is used for the event rate estimate. The magnitude distribution of the glints associated with catalogued objects is shown with the orange line. (Bottom) The distributions of the glints with single and multiple detections are shown with the black and green lines, respectively.
    Note that magnitudes are given as measured in 0.5 sec image.}
    \label{fig:dist_mag}
\end{figure}

The brightness of the detected glints ranges from 11--16 mag.
Figure \ref{fig:dist_mag} shows the distribution of magnitudes
for the detected glints as measured in 0.5 sec images (blue lines).
The number of glints increases
toward fainter magnitudes down to about 16 mag.
As shown in Figure \ref{fig:tpr}, the detection efficiency 
starts to drop around 1 magnitude brighter than the limiting magnitude.
Thus, the decrease in the number of glints on the fainter side is due to the
low detection efficiency
(a typical limiting magnitude is $\sim 16.5$ mag).
The glints associated with known satellites and debris
are found to be brighter ($<$ 14 mag), as shown in the orange line in the top panel.
In contrast, many fainter glints are not associated with
catalogued objects.
This is consistent with the results of the survey for GEO objects
in the past \citep{schildknecht07}:
most of the objects at the GEO with $> 15$ mag is not catalogued.

The number of uncatalogued objects in the orbit is important
  to understand the current space environment,
  although it is difficult to give an accurate estimate from our observations.
  In our observations, 70 catalogued objects produce 563 second-timescale
  glints.
  By roughly assuming that the frequency of producing second-timescale glints
  is the same for catalogued and uncatalogued objects,
  detection of 991 glints that are not associated with catalogued objects
  suggests that an order of 100 uncatalogued objects were detected
  in our observations.
  As the frequency to produce short-timescale glints may be smaller
  for the fainter objects
  (see Section \ref{sec:origin}), this number should be
  regarded as a lower limit.

\section{Discussion}
\label{sec:discussion}

\subsection{Origin of the glints}
\label{sec:origin}

As shown in Section \ref{sec:results}, 
most of the second-timescale glints detected with our observations
are likely to be caused by artificial satellites or space debris
at the high altitude orbits (such as GEO or HEO), reflecting the sunlight.
Below we discuss properties and characteristics of the objects
causing second-timescale glints.

In this work, we selected glints that appear as point sources.
At the GEO, for example, a typical apparent motion is 15 arcsec/sec.
Thus, for the objects to appear as point sources within
a seeing size ($\sim 4$ arcsec in our observations),
the intrinsic duration of the glints should be $\Delta t <$ 0.3 sec,
which is somewhat shorter than our exposure time of 0.5 sec.
This is consistent with the results obtained by
time-resolved observations with W-FAST \citep{nir21}.
Note that some glints also show faster apparent motion
  (Figure \ref{fig:motion}).
  For glints with 150 arcsec/sec, for example,
  the intrinsic duration should be even shorter ($\Delta t <$ 0.03 sec)
  to be detected as point sources.

We approximately estimate the size of the objects.
The observed flux from an object reflecting sunlight is expressed by considering
the solid angle of the object (e.g., \citealt{africano05,silha20}):
\begin{equation}
  f = \frac{\pi l^2}{r^2} A \psi(\phi) f_{\odot},
\end{equation}
where $l$ is the diameter of the object, $r$ is the distance from the observer,
$A$ is an albedo, and $f_{\odot}$ is the observed flux from the Sun.
The equivalent equation for magnitude is
\begin{equation}
  m = -2.5 \log (\pi A l^2 \psi (\phi)) + 5 \log (r) - m_{\odot},
\end{equation}
where $m_{\odot}$ is the visual magnitude of the Sun.
Here, $\psi (\phi)$ is a phase function, which defines
the angular dependence of the scattered light from the surface
($\phi$ is a phase angle of the reflection).

Typically, diffuse reflection is assumed to estimate the size of
satellites or space debris from optical observations \citep{africano05,schildknecht07,silha20},
which is known to to be a sound approximation.
In this case, the phase function of a Lambertian sphere is often adopted:
\begin{equation}
  \psi (\phi) = \frac{2}{3\pi^2} [ (\pi - \phi) \cos \phi + \sin \phi ].
\end{equation}
The phase function has a peak of $\psi = $ 0.2
when $\phi = 0$ (the Sun and the observer are in the same direction from the object).
For $\phi = 90$ deg, the phase function is $\psi = 0.07$.

\begin{figure}
	\includegraphics[width=\columnwidth]{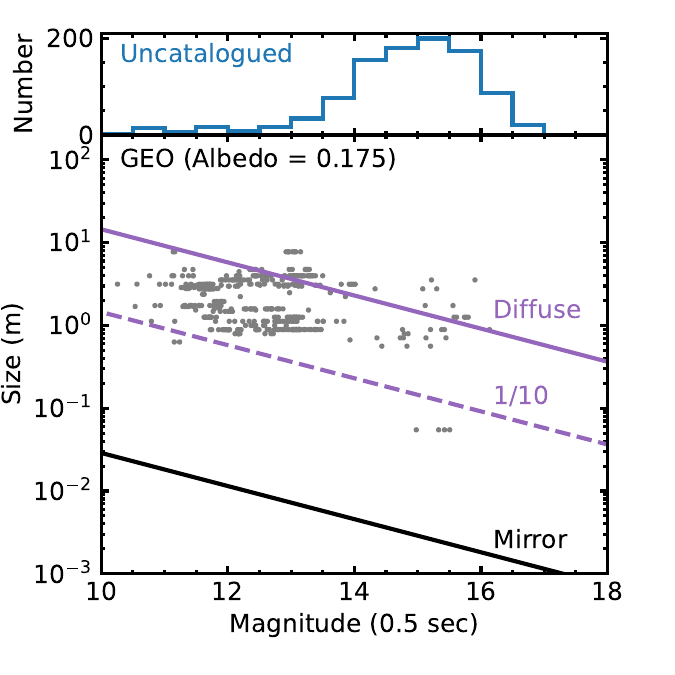}
        \caption{
          (Top) Brightness distribution for the glints that are not matched with catalogued objects.
          (Bottom) Relation between magnitude and size of the objects.
          The purple solid line shows the expected size assuming albedo of $A=0.175$ and
          scattered reflection with the phase angle of $\phi = 90$ deg at the GEO.
          The purple dashed line shows a smaller size by a factor of 10 to give a possible range.
          The black line shows the expected size assuming a perfect mirror reflection with albedo of $A=0.175$.
          The gray dots show the size estimated from radar cross section ($l = \sqrt{\rm RCS}$) for the glints
          associated with catalogued objects with an apogee of $> 30,000$ km.
        }
    \label{fig:size_mag}
\end{figure}

Figure \ref{fig:size_mag} shows the relation between size and observed magnitude
for objects at the GEO distance.
The albedo is assumed to be $A=$ 0.175, following the convention in
optical observations \citep{africano05,schildknecht07,silha20}.
The purple line shows the expected relation from scattered reflection
assuming a phase angle of $\phi = 90$ deg
(as most of glints are detected at this angle; Figure \ref{fig:dist_sep}).

Since we detect objects as short-timescale glints, 
true reflection of the objects may have a stronger angular dependence
than that of a Lambertian sphere.
In this case, the phase function becomes larger, which gives a smaller size estimate for a given brightness.
An extreme case is a perfect plane-mirror reflection 
as assumed in some previous works (e.g., \citealt{schaefer87,nir21,loeb24}).
Under this assumption, the phase function corresponds to 
$\psi = 1/\Omega_{\odot} = 1.7 \times 10^{4}$
only when the observer is located within the reflected sunlight
($\Omega_{\odot}$ is the solid angle of the Sun).
Due to the very high efficiency of the reflection,
this yields a substantially small size estimate for a given brightness
(black line in Figure \ref{fig:size_mag}).

The gray dots in Figure \ref{fig:size_mag} show the brightness of glints associated with known objects.
For these objects, we plot the size by assuming $l = \sqrt{\rm RCS}$,
where RCS is the radar cross section
obtained from SATCAT data in CelesTrak\footnote{\url{https://celestrak.org}}.
Although it is not necessarily true that the size estimated by radar
corresponds to that estimated from optical observations,
the RCS sizes lie between $0.1-1$ times the sizes estimated from a Lambertian sphere.

It should be noted that, in the comparison above, 
we do {\it not} correct the exposure time for the
observed brightness,
i.e., the brightness of the glints is assumed to be constant
over a 0.5 sec exposure.
This is not necessarily true as the typical duration is probably
somewhat shorter --
about 0.2--0.3 sec as estimated by \citet{nir21} or inferred
from the point-source shape (see above).
If the intrinsic duration is shorter than this,
the actual instantaneous brightness
would be brighter, which would lead to a larger size estimate.

Keeping this caveat in mind, the brightness of the faint glints
that are not associated with catalogued objects ($\sim 15-16$ mag,
the top panel of Figure \ref{fig:size_mag})
implies that they are caused by the objects with about 0.3-1 m in size.
Interestingly, past surveys for GEO objects show that the brightness distribution
has two distinct populations: one peaks around 12 mag (mainly catalogued objects)
and the other population (mainly uncatalogued objects) appears
around 15 mag or fainter \citep{schildknecht07}.
The fainter glints detected in our survey may correspond to
this fainter population
\footnote{
Another caveat on this possible correspondence is the size estimates
from dedicated surveys of the GEO objects.
Such surveys often detect the objects
by observing them as static source, loosing information of time variability.
Thus, if the objects show a large variability with a short timescale as shown in this paper, the size estimate may also be significantly affected.
}.

The origin of this fainter population is actively discussed.
In fact, the area-to-mass ratio of these faint objects
  is estimated to be $\sim 1-10 \ {\rm m^2 \ kg^{-1}}$
  from their orbital evolutions (through the effects of solar radiation pressure).
  This is substantially higher than that of catalogued objects
($\sim 0.01 \ {\rm m^2 \ kg^{-1}}$, \citealt{schildknecht07,molotov09}).
For comparison, a piece of paper has about $15 \ {\rm m^2 \ kg^{-1}}$.
The large area to mass ratio implies that these faint populations have plane- or sheet-like shapes,
e.g., multi-layer insulation pieces separated from spacecraft (e.g., \citealt{liou05}).
Interestingly, in this brightness range (15--16 mag),
the fraction of single glints becomes larger (Figure \ref{fig:dist_mag}).
This may be partly due to the difficulty of detection:
as it is close to the limiting magnitude, it is more challenging to identify
multiple glints.
Alternatively, the ``light curve'' of these faint objects
may differ from that of catalogued objects.
For example, they may show a larger-amplitude variability
or rarer ``flare'' events due to their sheet-like shapes.
Although estimating the object shape is beyond our scope in this paper,
wide-field video-mode observations may help to study the nature of this debris.

\begin{figure*}
\begin{center}  
\includegraphics[width=15cm]{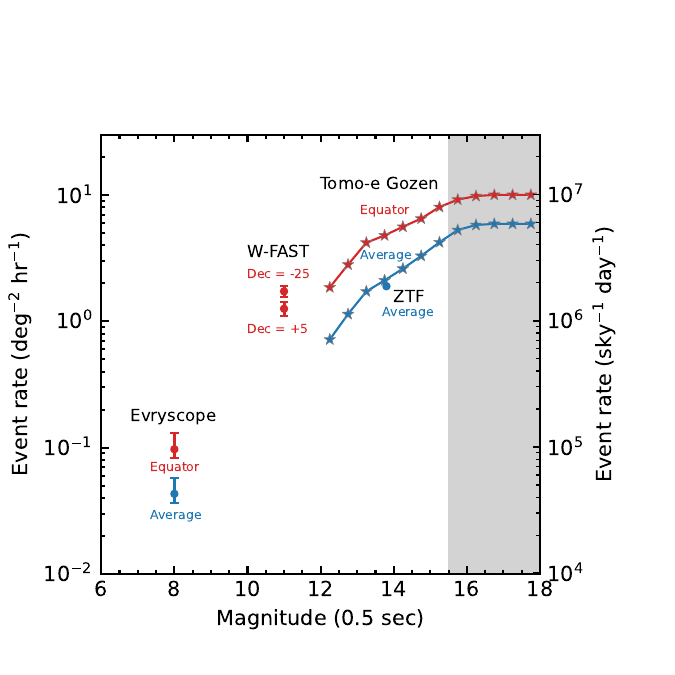}  
\caption{Cumulative areal event rate of second-timescale glints as a function of magnitude (0.5 sec equivalent).
  For our survey with Tomo-e Gozen, the average rate (blue) and the rate near the equator
  (red, $-20$ deg $<$ Dec $< +20$ deg) are shown.
Event rates estimated with Evryscope \citep{corbett20}, W-FAST \citep{nir21}, and ZTF \citep{karpov23} are also shown for comparison (see Table \ref{tab:rate}).
The magnitude range for Evryscope (120 sec exposure time) and ZTF (30 sec exposure time) is converted to 0.5 sec equivalent magnitude.
The magnitude range for W-FAST (0.04 exposure time, resolving the glint light curves) is also converted by assuming a typical intrinsic duration of the glints $\Delta t = 0.2$ sec.
The ZTF rate shown in this figure is the rate for all the candidates including
  uncatalogued sources.
To estimate the foreground rate for the future surveys, all the event rates in this figure are defined
as a detection-wise way:
multiple glints from the same objects are counted multiple times.
}
    \label{fig:rate}
\end{center}  
\end{figure*}

\subsection{Event rate}
\label{sec:rate}

Based on our observations, we estimate the areal event rate of
the second-timescale glints that appear as point sources.
The areal event rate is defined as $R = N_g / \eta S$,
where $N_g$ is the number of glints,
$\eta$ is the TPR of glint detection,
and $S$ is the areal exposure of the survey.
To avoid a large efficiency correction for the fainter objects
close to the limiting magnitude,
we only use glints detected in images deeper than 16.5 mag
(the areal exposure is 173 ${\rm deg^2 \ hr}$).
In this case, the detection efficiency falls below 0.5
around 15.5 mag.
Thus, we only count glints brighter than 15.5 mag.
For the detection efficiency,
we simply adopt $\eta = 0.9$ from the TPR of the SSD
(score $>$ 0.70 in Figure \ref{fig:tpr}).
As our purpose is understanding the foreground rate in transient surveys,
the event rate is measured in a detection-wise manner, i.e.,
multiple glints from the same objects are included in $N_g$ above.

Figure \ref{fig:rate} shows the cumulative areal event rate of
the detected glints as a function of magnitude (blue).
The estimated cumulative areal event rate of the glints 
is $4.7\pm 0.2$ ${\rm deg^{-2}\ hr^{-1}}$ at 15.5 mag.
The red points show the event rate near the equator ($-20$ deg $<$ Dec $<$ $+20$ deg),
reaching $9.0\pm 0.4$ ${\rm deg^{-2}\ hr^{-1}}$.
The event rate near the equator is higher than the average rate by a factor of about 2.

Our results show that the event rate of faint glints 
is higher than that of brighter glints.
In Figure \ref{fig:rate}, we also show the estimates of the areal rates
by \citet{corbett20}, \citet{nir21}, and \citet{karpov23}.
\citet{corbett20} performed observations with 120 sec exposure using Evryscope,
and detected glints with magnitudes down to 14 mag,
which corresponds to about 8 mag in a 0.5 sec exposure image.
\citet{nir21} performed observations with a 0.04 sec exposure (W-FAST),
and detected glints with 9--11 mag by resolving their light curves.
Assuming the intrinsic duration of the glints to be $\Delta t=0.2$ sec,
this corresponds to 10--12 mag in 0.5 sec exposure image.
\citet{karpov23} used ZTF images with 30 sec exposure time.
They estimated an intrinsic duration of $\Delta t=0.1$ sec from the point-source shape.
The Faintest ``instantaneous'' magnitude of the glints is about 12 mag in 0.1 sec, which corresponds to 13.8 mag in 0.5 sec exposure image.

The event rate of glints at 15.5 mag (in 0.5 sec) is higher than
that at about 11 mag and 8 mag by factors of about 5 and 100, respectively.
The event rate around 14 mag is intriguingly similar to that estimated
by \citet{karpov23}.
It should be cautioned that the detection algorithm, selection criteria
(definition of point sources),
and survey orientation (in terms of RA and Dec as well as solar separation angles)
are very different among these surveys,
which makes quantitative comparison challenging.
Nevertheless, it seems robust that fainter glints are more frequent,
as naturally expected from the larger number of fainter (and smaller) objects
found by past GEO surveys (e.g., \citealt{schildknecht07}).

\subsection{Implications for future optical surveys}

Finally, we discuss implications of our results for
future time-domain surveys as well as deep imaging surveys.
Our results and previous studies \citep{corbett20,nir21,karpov23}
demonstrate that optical time-domain surveys for second-timescale transients
suffer from a large number of foreground glints from satellites and space debris
(Figure \ref{fig:rate}).
For example, a sky rate of FRBs is estimated to be 
about $800 \ {\rm sky^{-1}\ day^{-1}}$, which corresponds to $8 \times 10^{-4} \ {\rm deg^{-2} \ hr^{-1}}$
\citep{chime21}.
The event rate of glints from satellites and debris is about 4 orders of magnitude
higher than the sky rate of FRBs.

One strategy to study second-timescale optical transients
  while avoiding frequent satellite/debris glints 
  is to use time and spatial coincidence with multi-wavelength observations.
For example, simultaneous optical observations of GRBs or FRBs
do not suffer from confusion with satellite glints.
According to our results, when GRBs or FRBs are well localized ($\ll 1$ deg$^2$),
the chance probability of detecting unrelated glints
at the same time as GRBs or FRBs is sufficiently small.

To search for second-timescale transients emitting only optical light,
a good strategy to avoid foreground objects is to perform observations
toward the Earth's shadow.
In fact, some video-mode surveys with Tomo-e Gozen have been conducted
by pointing to the small region corresponding to the Earth’s shadow
at the GEO altitude.
From these observations, upper limits on the event rate of second-timescale glints have been obtained:
$\le 4 \times 10^{-2} \ {\rm deg^{-2}\ hr^{-1}}$
by \citet{richmond20}
and $\le 4 \times 10^{-3} \ {\rm deg^{-2}\ hr^{-1}}$ by 
\citet{arima25}, excluding one potential candidate.
These limits are much lower than the event rate of the observed glints,
showing that surveys towards the Earth's shadow enable the study of
the short-timescale astrophysical transients.

These glints can be contaminants
not only for short-timescale transient searches, 
but also for deep observations with long exposures
using large-aperture telescopes \citep{loeb24,tyson24}.
Even though the duration of the glints is on the order of 0.5 sec,
the glints can remain in long-exposure images \citep{corbett20,karpov16}.
This effect can be significant for the deep surveys such as
the Legacy Survey of Space and Time (LSST) of the Vera C. Rubin Observatory.

For example, a population of glints with 16 mag in 0.5 sec exposure
image corresponds to 20.5 mag in 30 sec exposures,
which is easily detected in the LSST images.
Also, our results, combined with searches for fainter debris (e.g., \citealt{seitzer16,blake21}),
suggest that the event rate for fainter objects can be even higher.
For the depth of single exposure with the LSST ($\simeq 24$ mag),
  short-timescale glints with 19.5 mag in 0.5 sec can still be detected.
  If we assume that the event rate for fainter glints continues to increase
  with the same slope (Figure \ref{fig:rate}),
  the detection rate in LSST images can be an order of
  100 $\rm{deg^{-2} \ hr^{-1}}$.

Note that many of these glints may be recognized as elongated sources.
Assuming the apparent motion of 15 arcsec/sec as expected for objects
at the GEO,
glints with an intrinsic duration longer than 0.05 sec
should show a major axis of $>$ 0.75 arcsec,
which can be resolved in the LSST images.
Thus, if the intrinsic duration of the glints is about $0.1--0.3$ sec
as shown by \citet{nir21}, 
many of the glints should appear elongated
in long-exposure images.
Nevertheless, the intrinsic duration of the glints is not entirely clear
because of the unknown shape, rotation period, and surface properties.
Thus, there might be a population of objects with shorter intrinsic durations
that can appear as point sources.
\citet{tyson24} pointed out that, for the objects at the LEO,
the glints are out of focus for the LSST camera,
and signal-to-noise ratios for the satellite glints are significantly
  lower than those of point sources.
  The effect of defocus would be smaller for the objects at
  a higher altitude.
Thus, single-exposure LSST images may still suffer from
a relatively frequent contamination by satellite/debris glints.

\section{Summary} 
\label{sec:summary}

To search for second-timescale optical glints,
we performed wide-field video-mode observations
with 2 frames per second
by using Tomo-e Gozen mounted on the 1.05 m Kiso Schmidt telescope.
Our survey was conducted within 3 hours after the sunset,
mainly toward sky areas with solar separation angles around 90 deg.
The areal exposure of our survey is 290 deg$^2$ hr in total,
and 173 deg$^2$ hr for the data deeper than 16.5 mag.

By using the SSD method, we have identified 1554 point-source glints
that appear in only one time frame (0.5 sec).
Their brightness ranges from 11 mag to 16 mag, with a larger number of fainter glints.
Among them, 1088 glints are detected multiple times at different positions in the same chip
within 18 frames, while the remaining 466 glints are identified only once.
A majority of the glints brighter than 14 mag match with known satellites
and space debris near the equator.
The apparent motion of the glints with multiple detection
is broadly consistent with the objects in the GEO,
suggesting that the glints near the equator
  are mainly caused by objects in the GEO.
  We also identify many glints at high latitudes.
  There is a larger scatter in their apparent motions,
  and this population may be caused by objects in the
  HEO with a high inclination.

Majority of glints with 14--16 mag do not match with catalogued objects.
The estimated size for these objects is 0.3--1 m.
Such a population has already been recognized by past survey of objects at the GEO
\citep{schildknecht07}: they are likely to have sheet-like shapes,
such as insulation pieces separated from spacecraft.
Our results imply that on the order of 100 such objects may be
orbiting and they can cause significant foreground contamination
in searches for optical transients with a second timescale.

The event rate of second-timescale glints is estimated to be $4.7\pm 0.2$ ${\rm deg^{-2}\ hr^{-1}}$ (average)
and $9.0\pm 0.4$ ${\rm deg^{-2}\ hr^{-1}}$ (equator) at 15.5 mag,
which is significantly higher than that of brighter glints (Figure \ref{fig:rate}).
Faint second-timescale glints detected in our survey can still appear as 20.5 mag sources in 30 sec exposure images.
In addition, our results together with past surveys for the smaller objects suggest that 
the event rate for fainter objects can be even higher.
Thus, the deep survey with Rubin/LSST would detect many of these glints in
single-exposure images.

\begin{acknowledgments}

We thank Yasuhiro Imoto for his contributions to the development of the detection software. This work is supported by JST AIP Acceleration Research Grant Number JP20317829 and the Grant-in-Aid for Scientific research from JSPS (grant No. 24H00027).

\end{acknowledgments}





%
\facilities{Kiso Schmidt (Tomo-e Gozen)}

\software{astrometry.net \citep{lang10},
  astropy \citep{astropy13,astropy18,astropy22},  
  SExtractor \citep{bertin96},
  Skyfield \citep{rhodes19}
          }



\appendix


\begin{figure}[t]
  \includegraphics[width=\columnwidth]{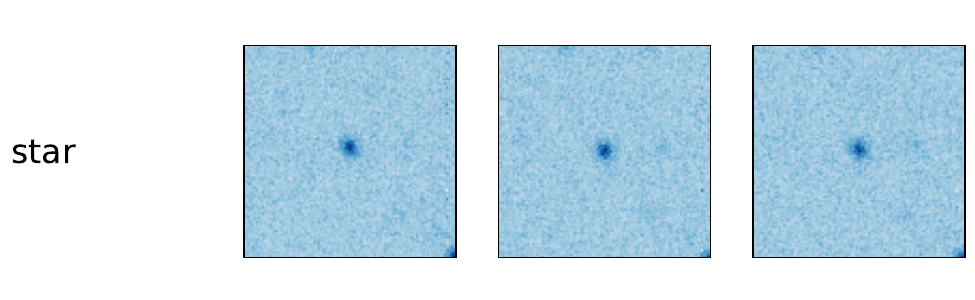}
  \includegraphics[width=\columnwidth]{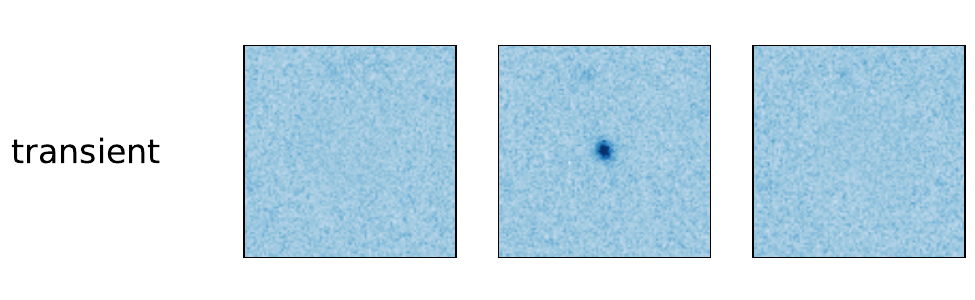}
  \includegraphics[width=\columnwidth]{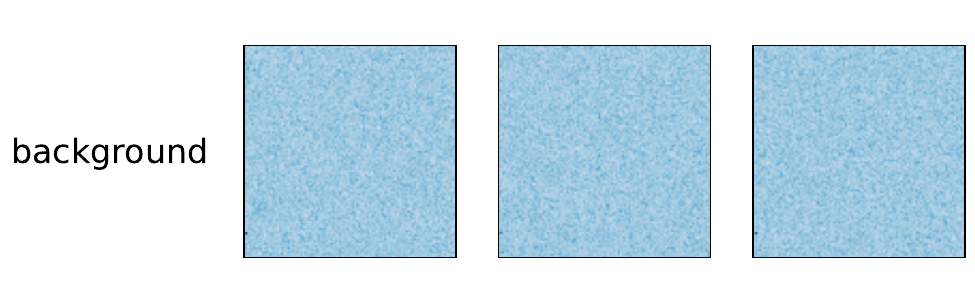}  
    \caption{Example of training data for the SSD: stationary source (top), transient (middle), and background (bottom). Three images for each category show three consequtive time frames.}
    \label{fig:ssd}
\end{figure}

\begin{figure}[t]
  \includegraphics[width=\columnwidth]{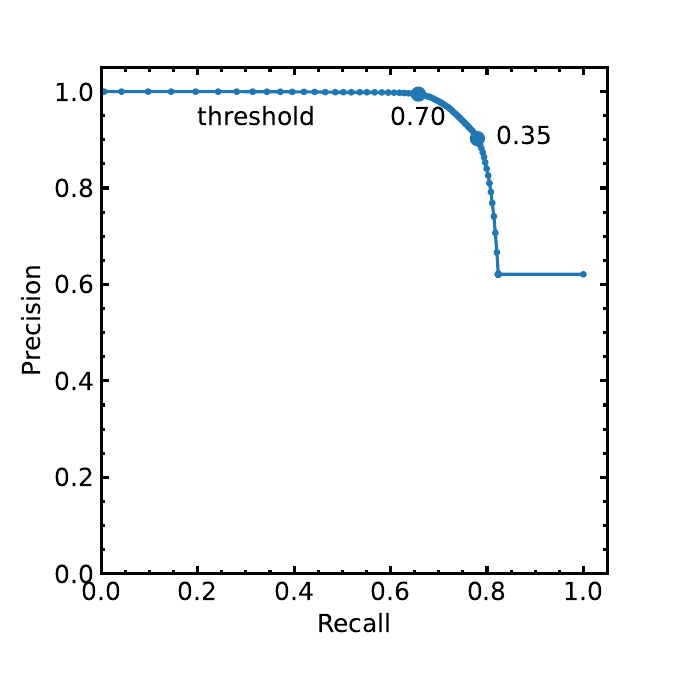}
  \caption{Precision-recall curve of our detection algorithm.
    Two large dots show the precision and recall with a threshold score of
    0.35 and 0.70. 
    Small dots are given with an interval of 0.01 in the threshold score.
      A jump in a high recall with the lowest threshold score is due to
      the very faint artificial sources which is always difficult to detect
      (see Figure \ref{fig:tpr}).}
    \label{fig:prc}
\end{figure}

\section{Detection Method Based on Single Shot Multibox Detector}

We have developed a dedicated detection method to identify
second-timescale glints based on Single Shot Multibox Detector
(SSD; \citealt{liu15}).
Our original movie data have dimensions of
2000 $\times$ 1128 pixels $\times$ 18 time frames.
For the input to the SSD, we use three consecutive time frames
(i.e., 2000 $\times$ 1128 pixels $\times$ 3 time frames).
Each image is normalized before being processed by the SSD.
In addition, we use three mask images at the corresponding time frames,
which highlight only the pixels with $>5 \sigma$ values.
In total, images with 6 channels are used as input.
Since the SSD accepts a patch of 300 $\times$ 300 pixels for each channel,
we subdivide the original image into 7 $\times$ 3 patches.
Thus, the input dimension for each patch is 300 $\times$ 300 $\times$ 6.
Detection and classification are performed for each patch.

For the base network of the SSD, we adopt a model structure similar to
so-called ``VGG16'' model \citep{simonyan14}.
Each block consists of 1--3 convolutional layers with $3 \times 3$ kernels
and $2 \times 2$ max pooling.
We set six blocks with dimensions of 
(150, 150, 18), (75, 75, 32), (38, 38, 64), (19, 19, 64),
(10, 10, 64), and (5, 5, 32).
Compared with the original structure of the SSD \citep{liu15},
  our model includes layers with $150 \times 150$ and $75 \times 75$
  pixels to detect small-scale objects (i.e., stars).
  On the other hand, we omit layers with $3 \times 3$ and $1 \times 1$ pixels
  (which are sensitive to large objects)
  since our targets are point sources.

For the SSD model prediction,
each block has pre-defined bounding boxes covering each pixel 
(i.e., default bounding boxes).
The SSD performs detection and classification for all the default bounding boxes.
For the detection, each default bounding box provides offsets to the true position,
yielding a bounding box enclosing the object (i.e., detection bounding box).
For each detection bounding box, classification probability is also estimated.
In this way, the SSD is able to detect objects of various sizes and
to classify them in a single step.
Finally, for each class, information from detection bounding boxes with a high probability
is gathered, and duplicated bounding boxes are removed
by non-maximum suppression.

The training dataset, i.e., images with ground-truth bounding boxes
and classification labels, was prepared from actual data of Tomo-e Gozen.
For the training data for the background and stationary objects,
we simply use the background region and object region, respectively.
For transient objects, we created the base of the training data
by artificially injecting point sources into 3490 movie datasets.
During the training procedure,
data augmentation is dynamically performed
by clipping, flipping, and rotating the images,
so that 6400 samples are used in every epoch of the training.
Training is performed for 100 epochs.
In each epoch, different augmented data are used,
which helps optimize the performance by avoiding overfitting.
For the loss function,
the Huber loss is adopted for the detection (i.e., localization loss)
while categorical cross-entropy is adopted for classification
(i.e., confidence loss).

Figure \ref{fig:prc} shows a precision-recall curve for our detection algorithm.
  The F1 score ($= 2$ (precision $\times$ recall)/(precision $+$ recall))
  is 0.84 and 0.79 at threshold scores of 0.35 and 0.70,
  respectively, demonstrating good performance of our detection algorithm.
  It is noted that the performance is measured with
  test data containing 20 artificial point sources per movie.
  The detection results consist of about 45,000 artificial sources
  (true positives) and 27,000 normal stars (false positives),
  yielding a precision of $\sim$ 0.6 even at a low threshold.
  Since the true fraction of second-timescale glints is smaller
  than that in the test data, the precision in actual observations would be lower.
  To reduce false positives in actual observations, 
  we adopt a conservative threshold score of 0.70 to
  maintain a relatively high recall while the precision is saturated.
  Note that the recall here is averaged over a wide magnitude range:
  the recall sharply drops toward lower signal-to-noise ratios
  as shown in Figure \ref{fig:tpr}.


\bibliographystyle{aasjournalv7}




\end{document}